\documentclass[floatfix,reprint, onecolumn, showpacs,nofootinbib]{revtex4-1}
\usepackage{natbib}
\usepackage{times}
\usepackage{amssymb,amsbsy,amsmath,amsfonts}
\usepackage{graphicx}
\usepackage{float}
\usepackage{morefloats}
\usepackage{rotating}
\usepackage{srcltx}
\newcommand{\eq}{\begin{eqnarray}}
\newcommand{\en}{\end{eqnarray}}
\begin{document}

\title{Low-energy interactions  of Nambu-Goldstone bosons with $D$ mesons in covariant chiral perturbation theory}

\author{L.S. Geng,$^{1,2}$ N. Kaiser,$^{2}$ J. Martin-Camalich,$^{3}$  and W. Weise$^2$}
\affiliation{
$^1$School of Physics and Nuclear Energy Engineering, Beihang University,  Beijing 100191,  China\\
$^2$Physik Department, Technische Universit\" at M\"unchen, D-85747 Garching, Germany\\
$^3$Departamento de F\'{\i}sica Te\'orica and IFIC, Universidad de
Valencia-CSIC, E-46071 Spain\\}

 \begin{abstract}
We calculate the scattering lengths of Nambu-Goldstone bosons interacting with $D$ mesons in a covariant formulation of
chiral perturbation theory, which satisfies heavy-quark spin symmetry and
  analytical properties of loop amplitudes. We compare our results with previous studies performed using heavy meson chiral perturbation theory and show that recoil corrections are sizable in most cases.
\end{abstract}


\date{\today}

\maketitle

\section{Introduction}
In recent years, studies of charmonium and open-charm systems have witnessed a renaissance. This was mainly
led by the experimental discoveries of various new particles, either the still largely mysterious $X$, $Y$, $Z$ particles
or the new open-charm states. Many of these states cannot be easily understood in conventional
quark models without introducing new degrees of freedom in addition to their basic $q\bar{q}$ structure, notably multi-quark components, i.e. $qq\bar{q}\bar{q}$.

An interesting resonance in this context is the $D_{s0}^*(2317)$  with a mass of $2317.8\pm0.6$ MeV and a small width of several MeV ($\Gamma<3.8$ MeV)~\cite{PDG}.
The $D_{s0}^*(2317)$ was first observed by the BaBar collaboration in the inclusive $D_s^+\pi^0$ invariant mass distribution
from $e^+e^-$ annihilation data at energies near 10.6 GeV~\cite{Aubert:2003fg} and later confirmed by Belle~\cite{Krokovny:2003zq} and CLEO~\cite{Besson:2003cp}. The nature of this state
has been extensively discussed in the literature~\cite{Bardeen:2003kt,Nowak:2003ra,vanBeveren:2003kd,Dai:2003yg,Narison:2003td,Chen:2004dy,Szczepaniak:2003vy,Browder:2003fk,Cheng:2003kg,Barnes:2003dj,Kolomeitsev:2003ac,
Guo:2006fu,Gamermann:2006nm}. All studies seem to agree that the coupling of the $D_{s0}^*(2317)$ to the nearby $DK$ threshold cannot be ignored. From this perspective, it is particularly interesting
to note
that the $D_{s0}^*(2317)$ can be ``dynamically'' generated in coupled-channel unitary approaches with interaction kernels provided by either chiral perturbation theory ($\chi$PT)~\cite{Kolomeitsev:2003ac,
Guo:2006fu} or a SU(4) Lagrangian ~\cite{Gamermann:2006nm}. Such approaches have provided many interesting results in the past few years, e.g.,
in explaining the nature of some low-lying hadronic states such as the $f_0(600)$ and the $\Lambda(1405)$ (for a comprehensive list of references, see Ref.~\cite{GarciaRecio:2010ki}).
In this dynamical picture of the $D_{s0}^*(2317)$, the interactions of $DK$ and coupled channels play a decisive role.
A quantity that characterizes the strength of such an interaction at low energies is the scattering length. Although it cannot measured
directly given the short life time of the $D$ mesons, it can nevertheless be studied on the lattice~\cite{Flynn:2007ki,Liu:2008rza}. The $s$-wave $DK$ scattering lengths
have recently been computed by several authors~\cite{Guo:2009ct,Liu:2009uz}. In Ref.~\cite{Guo:2009ct}, the calculation was performed using a covariant formulation of
$\chi$PT up to next-to-leading-order (NLO) and its unitarized version. While in Ref.~\cite{Liu:2009uz}, the calculation was performed
using the Heavy-Meson $\chi$PT (HM$\chi$PT)~\cite{Wise:1992hn,Yan:1992gz,Burdman:1992gh,Jenkins:1992hx,Casalbuoni:1996pg} in the heavy quark limit up to next-to-next-to-leading (NNLO) order, where recoil effects have been
neglected. The authors cautioned, however, that since the $D$ mesons are not heavy enough, recoil corrections may not be small and have to be studied.

To our knowledge, recoil corrections have so far not been systematically studied in $\chi$PT describing the interactions between heavy-light mesons and
Nambu-Goldstone bosons. They have been, however, studied quite extensively in the one-baryon sector with three flavors $u$, $d$, and $s$.
There these corrections were found to be fairly large and play an important role in the studies of many physical observables~\cite{Geng:2008mf,Geng:2009ys,Geng:2009ik,MartinCamalich:2010fp}.
Because of the large baryon mass that does not vanish in the chiral limit,
covariant baryon $\chi$PT often faces  the so-called power-counting-breaking (PCB) problem~\cite{Gasser:1987rb}.
This problem has been traditionally dealt with using a dual expansion in terms of both $p/\Lambda_\chi$ and $1/M_B$, where $p$ is a generic small quantity, $M_B$ the baryon mass and $\Lambda_\chi=4\pi f_\pi$ the chiral symmetry breaking scale. This is the celebrated heavy baryon $\chi$PT ~\cite{Jenkins:1990jv,Bernard:1995dp}.  Though very successful in describing many observables, this approach is not
covariant and modifies the analyticity of loop amplitudes. More importantly, from a practical point of view it suffers from slow convergence, particularly in the 3-flavor sector. To overcome these problems, several other approaches to deal with
the PCB problem have been proposed, which among others include the infrared (IR)~\cite{Becher:1999he} and extended-on-mass-shell (EOMS)~\cite{Gegelia:1999gf,Fuchs:2003qc} renormalization schemes.
While the IR scheme was found to introduce artificial cuts in certain cases (see, e.g., Ref.~\cite{Holstein:2005db}), the EOMS approach is fully covariant and conserves the analytical properties of
loop amplitudes.
In a series of applications~\cite{Geng:2008mf,Geng:2009ys,Geng:2009ik,MartinCamalich:2010fp}, it has been shown that the EOMS approach also improves the convergence behavior of SU(3) baryon $\chi$PT.

The main purpose of the present work is to study the scattering lengths of Nambu-Goldstone bosons ($\phi$) interacting with $D$ mesons in a covariant formulation of
$\chi$PT by using the EOMS scheme to remove the PCB terms induced by the large $D$ meson masses. Given the fact that
the $D$ meson mass ($\sim1.9$ GeV) is much larger than the nucleon mass, it is anticipated that the recoil corrections should be smaller than those in the
nucleon case.
Nevertheless, such recoil corrections may still be sizable as we will show in this work.

This paper is organized as follows. In Section II, we introduce the relevant effective Lagrangians and explain briefly the
EOMS renormalization scheme. In Section III, we show the numerical results, compare them with those of earlier studies by paying special
attention to the recoil corrections, followed by a short summary in Section IV.

\section{Theoretical framework}
In this section we introduce the chiral Lagrangians relevant to the present study up to NNLO and explain briefly the EOMS renormalization scheme
used to remove the PCB pieces appearing in the covariant loop calculation.
\subsection{Chiral Lagrangians}
To introduce the chiral effective Lagrangians, one must specify a power counting rule.
In the present case the light meson masses $m_\phi$ and the field gradients $\partial_\mu\phi$ are counted as $\mathcal{O}(p)$, while
$\partial_\mu P$, $\partial_\nu P^*_\mu$, $m_P$, and $m_{P^*}$ are counted as $\mathcal{O}(1)$, where $\phi$ denotes
the Nambu-Goldstone bosons,
$P=(D^0,D^+,D^+_s)$ and $P^*_\mu=(D^{*0},D^{*+},D^{*+}_s)_\mu$ are the $D$ and $D^*$ meson fields.  The Nambu-Goldstone boson propagator,$\frac{i}{q^2-m_\phi^2}$,
is counted as $\mathcal{O}(p^{-2})$, while the heavy-light pseudoscalar and vector meson propagators, $\frac{i}{q^2-m_P^2}$ and $\frac{i}{q^2-m_{P^*}^2}(-g^{\mu\nu}+\frac{q^\mu q^\nu}{m_{P^*}^2})$, are counted as $\mathcal{O}(p^{-1})$.

The leading order chiral Lagrangian describing the self-interaction of Nambu-Goldstone bosons has the standard form:
\begin{equation}
 \mathcal{L}^{(2)}=\frac{1}{48 f_0^2}\langle ((\partial_\mu \Phi)\Phi-\Phi\partial_\mu\Phi)^2+\mathcal{M}\Phi^4\rangle,
\end{equation}
where $\Phi$ collects the pseudoscalar octet fields
\begin{equation}
 \Phi=\sqrt{2}\left(
\begin{array}{ccc}
 \frac{\pi^0}{\sqrt{2}}+\frac{\eta}{\sqrt{6}} & \pi^+ &K^+\\
  \pi^- & -\frac{\pi^0}{\sqrt{2}}+\frac{\eta}{\sqrt{6}} & K^0\\
  K^- & \bar{K}^0 & -\frac{2}{\sqrt{6}}\eta
\end{array}
\right),
\end{equation}
$\mathcal{M}=\mathrm{diag}(m_\pi^2,m_\pi^2,2 m^2_K-m_\pi^2)$ is the mass matrix, and $f_0$ is the pseudoscalar decay constant
in the chiral limit. Here and in the following, $\langle\cdots\rangle$ always denotes the trace in the respective flavor space.

The lowest order chiral Lagrangian for the heavy-light pseudoscalar and vector mesons is~\footnote{Because of heavy-quark spin symmetry, the pseudoscalar and vector mesons can be assigned to the same multiplet.}
\begin{eqnarray}\label{lag:lo}
 \mathcal{L}^{(1)}&=&\langle \mathcal{D}_\mu P \mathcal{D}^\mu P^\dagger\rangle-\mathring{m}_D^2\langle P P^\dagger\rangle-\langle\mathcal{D}_\mu P^{*\nu}\mathcal{D}^\mu P^{*\dagger}_\nu\rangle+\mathring{m}_D^2\langle P^{*\nu}P^{*\dagger}_\nu\rangle\nonumber\\
                  &&+ i g \langle P^*_\mu u^\mu P^\dagger -P u^\mu P^{*\dagger}_\mu\rangle
                  +\frac{g}{2\mathring{m}}\langle  (P^*_\mu u_\alpha \partial_\beta P^{*\dagger}_\nu-\partial_\beta P^*_\mu u_\alpha P^{*\dagger}_\nu)\epsilon^{\mu\nu\alpha\beta}\rangle
\end{eqnarray}
where $\mathcal{D}_\mu P_a=\partial_\mu P_a  - \Gamma_\mu^{ba} P_b$ and $\mathcal{D}^\mu P^\dagger_a=\partial^\mu P^\dagger_a+\Gamma^\mu_{ab} P^\dagger_b$ with $a$ ($b$) the SU(3)
flavor index, $g$ is the heavy-light pseudoscalar-vector coupling constant of dimension 1 and $\mathring{m}$ is the
mass of the heavy-light meson in the chiral limit. The vector and axial-vector currents,
$\Gamma_\mu$ and $u_\mu$, are defined as
\begin{equation}
 \Gamma_\mu=\frac{1}{2}(u^+\partial_\mu u+u\partial_\mu u^+)\quad\mbox{and}\quad
 u_\mu=i(u^\dagger\partial_\mu u-u\partial_\mu u^\dagger)
\end{equation}
with $u^2=U=\exp\left(\frac{i\Phi}{f}\right)$. The numerical value of $g$ can be fixed by reproducing the $D^{*+}\rightarrow D^0\pi^+$ decay width. Using the PDG average, $\Gamma_{D^{*+}}=96\pm22$ keV and $BR_{D^{*+}\rightarrow D^0\pi^+}=(67.7\pm0.5)\%$~\cite{PDG},
one obtains $\Gamma_{D^{*+}\rightarrow D^0\pi^+}=\frac{1}{12 \pi}\frac{g^2}{f_\pi^2}\frac{|q_\pi|^3}{m_{D^{*+}}^2}=65\pm15$ keV and accordingly $g=1177\pm 137\,\mathrm{MeV}$.

The NLO Lagrangian relevant to our study is~\footnote{$D^*$ mesons in NLO and NNLO Lagrangians do not contribute to the NNLO $D\phi$ scattering lengths and therefore will not be explicitly shown
in this work.}
\begin{equation}\label{lag:nlo}
\mathcal{L}^{(2)}=-2\left[ c_0\langle P P^\dagger \rangle \langle \chi_+\rangle
-c_1\langle P \chi_+ P^\dagger\rangle-c_2\langle PP^\dagger\rangle \langle u^\mu u_\mu\rangle
-c_3\langle P  u^\mu u_\mu P^\dagger\rangle \right],
\end{equation}
where
$\chi_\pm=u^\dagger \mathcal{M} u^\dagger\pm u\mathcal{M} u$.
Here we have adopted a convention consistent with that of Ref.~\cite{Liu:2009uz} for the
purpose of later comparison. In general, there are more terms contributing to $\phi P$ scattering, e.g.,
$\langle \partial_\mu P \partial_\nu P^\dagger\rangle \langle u^\mu u^\nu\rangle$, $\langle \partial_\mu P u^\mu u^\nu \partial_\nu P^\dagger\rangle$,
$\langle \partial_\mu P u^\nu u^\mu \partial_\nu P^\dagger\rangle$. However, at $\phi P$ threshold, it can be easily shown that these terms lead to the same structure as those
proportional to $c_2$ and $c_3$ and therefore can be neglected. The low-energy-constant (LEC) $c_1$ can be determined
from the mass splitting between strange and non-strange heavy-light mesons within the same doublet, i.e.,
\begin{equation}
-8 c_1 (m_K^2-m_\pi^2)=(m_{D_s}^2-m_D^2+m_{D_s^*}^2-m_{D^*}^2)/2.
\end{equation}
Using the masses given in Table~\ref{table:par}, we obtain $c_1=-0.225$.

\begin{table*}[htpb]
      \renewcommand{\arraystretch}{1.6}
     \setlength{\tabcolsep}{0.4cm}
     \centering
     \caption{\label{table:par}Numerical values of (isospin-averaged) masses~\cite{PDG} and decay constants~\cite{Amoros:2001cp} (in units of MeV) used in the present study.
The $f_K/f_\pi$ ratio is consistent with the latest determination~\cite{Rosner:2010ak} while the $f_\eta/f_\pi$ ratio is in agreement with that determined in a number of other approaches,
see, e.g., Ref.~\cite{Kolesar:2008fu}. The eta meson mass is calculated using the Gell-Mann-Okubo mass relation: $m_\eta^2=(4 m_K^2-m_\pi^2)/3$.}
     \begin{tabular}{cccccccccccc}
     \hline\hline
    $\mathring{m}_{D}$  &  $m_{D^*_s}$ & $m_{D^*}$ & $m_{D_s}$   & $m_D$  &  $m_\pi$ & $m_K$  &    $m_\eta$   &     $f_\pi$  &$f_K$   & $f_\eta$   \\
   $1972.1$    & 2112.3 & 2008.6  & 1968.5  & 1867.2 & 138.0  & 495.6 & 566.7 & 92.4  & $1.22 f_\pi$      & $1.31 f_\pi$\\
 \hline\hline
    \end{tabular} 
\end{table*}

At NNLO one has~\footnote{In Ref.~\cite{Liu:2009uz} only one such term, that proportional to $\kappa$,
was considered.}
\begin{eqnarray}\label{lag:nnlo}
 \mathcal{L}^{(3)}&=&-\frac{i}{2}\kappa \langle \partial^\mu P (x^-_\mu) P^\dagger- P(x^-_\mu)\partial^\mu P^\dagger\rangle +
\frac{\gamma_0}{2}\langle\partial^\mu P \Gamma_\mu P^\dagger-P \Gamma_\mu \partial^\mu P^\dagger\rangle\langle \chi_+\rangle\nonumber\\
                  &&+\gamma_1\langle \partial^\mu P \chi_+\Gamma^\mu P^\dagger-P\Gamma_\mu \chi_+\partial^\mu P^\dagger \rangle
                  +\gamma_2\langle \mathcal{D}^\mu P P^\dagger-P \mathcal{D}^\mu P^\dagger\rangle \langle \mathcal{D}_\mu \chi_+\rangle,
\end{eqnarray}
where
$x^-_\mu=[\chi_-,u_\mu]$. Although the number of relevant LECs at NNLO is considerably smaller than that present in
meson-baryon scattering processes~\cite{Oller:2006yh,Frink:2006hx}, it is still relatively large considering the scarcity of lattice data. Therefore in the present study we follow
 the approach adopted
in Refs.~\cite{Mai:2009ce,Liu:2006xja} in studies of meson-baryon scatterings and put the NNLO LECs to zero. This is acceptable in the present work because
we focus on the recoil corrections and not on the absolute values of the scattering lengths.

\subsection{Power counting restoration and the EOMS renormalization scheme}
In a covariant formulation of $\chi$PT describing the interactions between heavy-light mesons and Nambu-Goldstone bosons,
one has to face the PCB problem. That is to say, in the calculation of
 a loop diagram one may find terms with a chiral order lower than that determined by the naive
 power-counting as prescribed in the previous sub-section. Such analytical PCB terms can be removed, just as
in baryon $\chi$PT, by using the heavy-meson expansion, the IR or the EOMS renormalization prescriptions. The essence of the EOMS approach lies in
the fact that $\chi$PT, by construction, contains all the structures allowed by symmetry. Therefore, the PCB pieces appearing
in a loop calculation can always be removed by redefining the corresponding LECs.
This is equivalent to removing the finite PCB pieces directly from the loop results.
In practice, this can be achieved in two slightly different ways: 1) one can first perform the loop calculation analytically, and then remove
the PCB terms, or 2) one can first perform an expansion in terms of the inverse heavy-meson mass, $1/m_H$, calculate the PCB terms and then subtract them from
the full results. It should be noticed that the second prescription is different from the heavy-meson (baryon) expansion because
in general integration and expansion may not commute. But since the PCB terms are finite and analytical, the second prescription should always work.

In the present study, we have explicitly checked that all the PCB terms appearing in our loop calculation can be removed by redefining the LECs introduced in the
previous sub-section.
\begin{figure}[t]
\centerline{\includegraphics[scale=0.5]{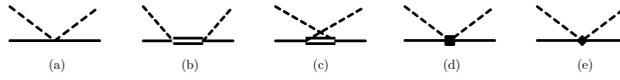}}
\caption{Tree-level contributions at LO [(a),(b),(c)], NLO (d) and NNLO (e).\label{fig1}}
\end{figure}

\begin{figure}[h]
\centerline{\includegraphics[scale=0.5]{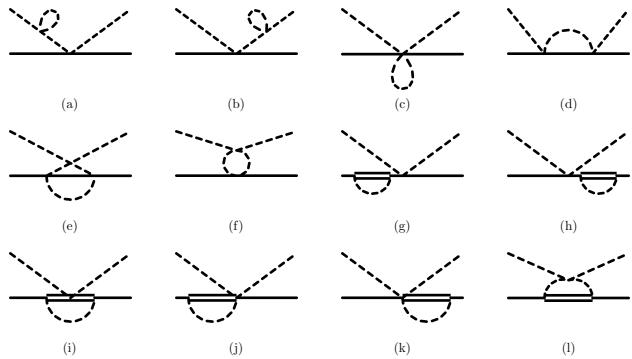}}
\caption{NNLO loop contributions that survive in the infinite heavy-meson mass ($m_H\rightarrow\infty$) limit.\label{fig2}}
\end{figure}
\begin{figure}[h]
\centerline{\includegraphics[scale=0.5]{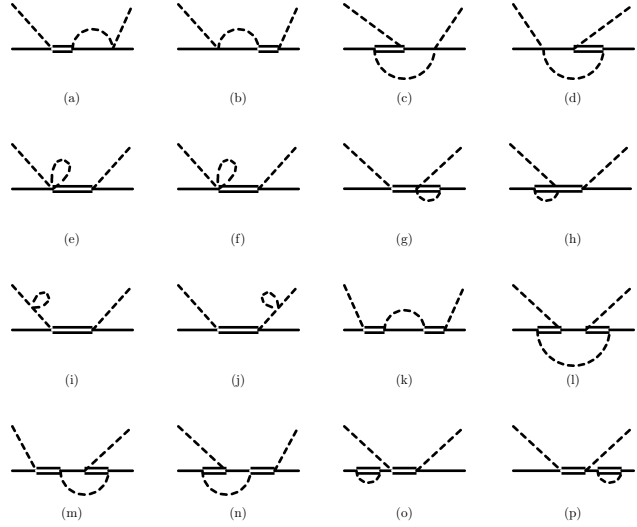}}
\caption{NNLO loop contributions that vanish in the $m_H\rightarrow\infty$ limit. Their crossed counterparts are not shown here
but included in the calculation.\label{fig3}}
\end{figure}
\section{Results and discussion}
The tree-level diagrams contributing to the (dimensionless) threshold $T$-matrices at LO [Fig.~(1a,1b,1c)], NLO [Fig.~(1d)] and NNLO [Fig.~(1e)] are shown in Fig.~\ref{fig1}.
Among the three LO diagrams, the contact term [Fig.~(1a)] yields the following results
\begin{eqnarray}
T_1^{(1)}=-\frac{\mathring{m}_D m_K}{f_0^2},\quad T_2^{(1)}=0,\quad T_3^{(1)}=0,\quad T_4^{(1)}=\frac{2 \mathring{m}_D m_K}{f_0^2},\quad T_5^{(1)}=0,\quad T_6^{(1)}=-\frac{m_{\pi } \mathring{m}_D}{f_0^2},\\
T_7^{(1)}=\frac{2 m_{\pi }
   \mathring{m}_D}{f_0^2},\quad T_8^{(1)}=0,\quad T_9^{(1)}=\frac{\mathring{m}_D m_K}{f_0^2},\quad T_{10}^{(1)}=-\frac{\mathring{m}_D m_K}{f_0^2},\quad T_{11}^{(1)}=\frac{\mathring{m}_D m_K}{f_0^2},
\end{eqnarray}
where $1,\cdots 11$ denote the $D_s K$, $DK(1)$, $D_s\pi$, $DK(0)$, $D_s\eta$, $D\pi(3/2)$, $D\pi(1/2)$, $D\eta$, $D_s\bar{K}$, $D\bar{K}(1)$ and $D\bar{K}(0)$
channels, respectively. In labeling the 11 channels, we have explicitly shown their isospin in the parentheses whenever necessary. In the above results, $f_0$ is the Nambu-Goldstone boson decay constant in the chiral limit. The contributions given by diagrams $(b)$ and $(c)$ of Fig.~1 also count as
$\mathcal{O}(p)$, but at threshold they are in fact $\mathcal{O}(p^2)$ and have the same structure as those provided by the NLO Lagrangians and therefore can be effectively taken
into account by redefining the LECs $C_1$ and $C_0$ (see below). Hence we will not calculate them explicitly in the present work. Following
the same arguments, we will also neglect similar diagrams at NNLO with one of the LO vertices replaced by a NLO vertex.

Lagrangian (\ref{lag:nlo}) provides the NLO tree-level contributions [Fig.~(1d)]~\footnote{These results are the same as those
of Ref.~\cite{Liu:2009uz} except that there the expression for $T^{(2)}_4$ is incorrect~\cite{Liu:private}.}:
\begin{eqnarray}
T_1^{(2)}=\frac{C_1 m_K^2}{f_0^2},\quad
T_2^{(2)}=\frac{\left(C_0+C_1\right) m_K^2}{2 f_0^2},\quad
T_3^{(2)}=\frac{\left(C_0+C_1\right)m_{\pi }^2}{2 f_0^2},\quad
T_4^{(2)}=\frac{\left(3 C_1-C_0\right) m_K^2}{2 f_0^2},\\
T_5^{(2)}=\frac{\left(7 C_1-C_0\right) m_{\eta}^2-16 c_1 \left(m_{\eta }^2-m_{\pi }^2\right)}{6 f_0^2},\quad
T_6^{(2)}=\frac{C_1 m_{\pi }^2}{f_0^2},\quad
T_7^{(2)}=\frac{C_1 m_{\pi}^2}{f_0^2},\\
T_8^{(2)}=\frac{4 c_1 \left(m_{\eta }^2-m_{\pi }^2\right)+\left(C_0+2 C_1\right) m_{\eta }^2}{3f_0^2},\quad
T_9^{(2)}=\frac{C_1 m_K^2}{f_0^2},\quad
T_{10}^{(2)}=\frac{C_1 m_K^2}{f_0^2},\quad
T_{11}^{(2)}=\frac{C_0 m_K^2}{f_0^2},
\end{eqnarray}
where we have introduced two combinations of the 4 LECs: $C_1=4 (2 c_0-c_1+2 c_2+c_3)$ and $C_0=4 (2 c_0 + c_1 + 2 c_2 - c_3)$.

The NNLO loop contributions can be separated into two groups: those that survive in the infinite heavy-meson mass ($m_H\rightarrow \infty$) limit (Fig.~\ref{fig2}) and
those that vanish in the $m_H\rightarrow\infty$ limit (Fig.~\ref{fig3}).
For the first group, our results recover those of Ref.~\cite{Liu:2009uz} in the $m_H\rightarrow\infty$ limit.
For the second group, our calculation shows that they indeed vanish in
the $m_H\rightarrow\infty$ limit but are not negligible in a covariant calculation, as shown below. It should be noted that all the loop contributions can be calculated analytically except the box-diagrams [Fig.(3l) and its crossed counterpart]. However, the analytical results are
quite involved and therefore we refrain from showing them explicitly. As explained earlier, in the present work we are going to neglect all
the NNLO counter terms. Accordingly we have removed from our loop results all the NNLO analytical terms. Therefore our loop results contain only NNLO non-analytical and higher-order terms. This way the differences between our covariant loop results and those of HM$\chi$PT are strictly recoil corrections.

The $s$-wave scattering lengths are related to the
threshold $T$-matrix elements through
\begin{equation}\label{eq:sl}
 a=\frac{1}{8\pi (m_\phi+m_P)}T_\mathrm{thr.}=\frac{\mathring{m}_D}{8\pi (m_\phi+m_P)}\tilde{T}_\mathrm{thr.},
\end{equation}
where $m_\phi$ is the Nambu-Goldstone mass induced by explicit chiral symmetry breaking and $m_P$ is the $D$ meson mass of a particular
channel. It should be noted that although in the calculation of $T_\mathrm{th}$ we have
used the average $D$ meson mass, in evaluating the scattering lengths [Eq.~(\ref{eq:sl})] we  use
the physical masses as given in Table \ref{table:par}. We also use the physical decay constants instead of the chiral limit ones since
the differences between them are of higher order.  Furthermore, we have introduced $\tilde{T}$ with dimension of a length for a more transparent comparison with
the HM$\chi$PT results of Ref.~\cite{Liu:2009uz}.

The unknown low-energy constants, $C_1$ and $C_0$, could  in principle be
fixed by reproducing data, which are however not yet available. As in Ref.~\cite{Liu:2009uz}, one can fix them
by reproducing the preliminary lattice QCD data (in units of fm): $a_{D\pi(3/2)}=-0.16(4)$, $a_{D\bar{K}(1)}=-0.23(4)$, $a_{D_s\pi}=0.00(1)$, and
$a_{D_s K}=-0.31(2)$.

In Table \ref{table2}, we list the $T$-matrix elements order by order ($\tilde{T}^{(1)}$ for LO and $\tilde{T}^{(2)}$ for NLO)
and the scattering lengths for the 11 independent (strangeness, isospin) channels computed in the $m_H\rightarrow \infty$ limit. Since we have neglected the
NNLO counter terms, we have denoted the NNLO results by $\tilde{T}^{(3)}_L$ stressing the fact that they contain only loop contributions.
We have fixed the LECs $C_1$ and $C_0$  by a least-squares fit of the four lattice points~\cite{Liu:2008rza}. We obtain
a $\chi^2/dof=5.0$ and the following values for the LECs: $C_1=0.4\pm1.2$ and $C_0=9.6\pm10.4$ at the $95\%$ confidence level.
Clearly, the four lattice data do not constrain well the two LECs.

In Table \ref{table3}, we show the values of the $T$-matrix elements and scattering lengths found in the covariant approach.
 The fit yields a $\chi^2/dof=4.5$ and the following values for the two LECs: $C_1=2.0\pm1.2$ and $C_0=4.0\pm10.4$.
Comparing the relativistic and non-relativistic results, one can easily see that
the recoil corrections are sizable.  For instance, $\tilde{T}^{(3)}_L$ for $D_s K$ changes from $-1.9$ fm to  $-5.1$ fm, and  $\tilde{T}^{(3)}_L$ for
$D_s \eta$ changes from  $0.1+9.7i$ fm to $-4.4+5.8i$ fm when going from HM$\chi$PT to covariant $\chi$PT.

In obtaining the numbers shown in Tables \ref{table2} and \ref{table3}, we have removed the ultraviolet divergences~\footnote{They appear in both 
HM$\chi$PT (see also Ref.~\cite{Liu:2009uz}) and in covariant $\chi$PT. The HM$\chi$PT results could have been made renormalization scale independent if we had kept the NNLO counter terms.} by the
modified minimal subtraction ($\widetilde{MS}$) renormalization scheme and have set the renormalization scale $\mu$ to $4\pi f_\pi$ . In the covariant $\chi$PT,
there are two heavy scales, $\Lambda_\chi$ and $\mathring{m}_D$. We have checked that our results would remain qualitatively the same if we
had set $\mu$ to $\mathring{m}_D$.

In Table \ref{table4}, the loop contributions $\tilde{T}^{(3)}_L$ calculated in covariant $\chi$PT are
decomposed into three parts: part A comes from diagrams Figs.~(2a-2f); part B comes from diagrams Figs.~(2g-2l); part C comes from diagrams
Fig.~3. It is clear that diagrams Figs.~(2a-2f) provide the most important contributions, while those from diagrams Fig.~3 are similar in size to those
from diagrams Figs.~(2g-2l), which is different in HM$\chi$PT, where the contributions from diagrams Fig.3 vanish.

\begin{table*}[htpb]
      \renewcommand{\arraystretch}{1.6}
     \setlength{\tabcolsep}{0.1cm}
     \centering
     \caption{Threshold $T$-matrix elements $\tilde{T}$ and scattering lengths $a$ (in units of fm) in the non-relativistic $\chi$PT up to NNLO ($\tilde{T}^{(1)}$
for LO, $\tilde{T}^{(2)}$ for NLO and $\tilde{T}^{(3)}$ for NNLO including only loop contributions).
The preliminary lattice QCD results~\cite{Liu:2008rza} are denoted by $a$(lQCD) and have been fitted to fix the two LECs $C_1$ and $C_0$.\label{table2}}
     \begin{tabular}{c|c|cc|cc|c|ccc|cc}
     \hline\hline
$(S,I)$    & $(2,1/2)$& \multicolumn{2}{|c|}{$(1,1)$}   &\multicolumn{2}{|c|}{$(1,0)$}  &$(0,3/2)$ & \multicolumn{3}{|c|}{$(0,1/2)$}   & $(-1,1)$  & $(-1,0)$\\
Channels   & $D_s K$& $D K$& $D_s \pi$ & $D K$ &$D_s \eta$ & $D \pi$ &$D\pi$  &$D\eta$ &$D_s\bar{K}$ & $D\bar{K}$ & $D\bar{K}$\\\hline
$\tilde{T}^{(1)}$  & $-7.7$ & 0 & 0 & 15.4 & 0 & $-3.2$ & 6.4 & 0 & 7.7 & $-7.7$ & $7.7$ \\
$\tilde{T}^{(2)}$  &  0.8 & 9.7 & 1.1 & $-8.0$ & $-1.2$ & 0.1 & 0.1 & 7.0 & 0.8 & 0.8 & 18.5 \\
$\tilde{T}^{(3)}_L$  & $-1.9$ & $-1.6+5.7 i$ & $-1.1$ & 3.5 & $0.1+9.7 i$ & $-0.8$ & 0.3 & $0.8+4.8 i$ &
   $0.2+8.5 i$ & $-3.4$ & $4.8$ \\
$a$ &  $-0.28$ & $0.27+0.19 i$ & 0.00 & 0.36 & $-0.04+0.30 i$ &
   $-0.15$ & 0.26 & $0.25+0.16 i$ & $0.28+0.27 i$ & $-0.34$ & $1.03$\\
$a$(lQCD)     & $-0.31(2)$ &  & $0.00(1)$&  &  &  $-0.16(4)$ & & & & $-0.23(4)$&\\

 \hline\hline
    \end{tabular} 
\end{table*}

\begin{table*}[htpb]
      \renewcommand{\arraystretch}{1.6}
     \setlength{\tabcolsep}{0.1cm}
     \centering
     \caption{Same as Table \ref{table3}, but with the threshold $T$-matrix elements and scattering lengths calculated in relativistic $\chi$PT.\label{table3}}
     \begin{tabular}{c|c|cc|cc|c|ccc|cc}
     \hline\hline
$(S,I)$  & $(2,1/2)$& \multicolumn{2}{|c|}{$(1,1)$}   &\multicolumn{2}{|c|}{$(1,0)$}  &$(0,3/2)$ & \multicolumn{3}{|c|}{$(0,1/2)$}   & $(-1,1)$  & $(-1,0)$\\
Channels & $D_s K$& $D K$& $D_s \pi$ & $D K$ &$D_s \eta$ & $D \pi$ &$D\pi$  &$D\eta$ &$D_s\bar{K}$ & $D\bar{K}$ & $D\bar{K}$\\\hline
$\tilde{T}^{(1)}$       &  $-7.7$ & 0 & 0 & 15.4 & 0 & $-3.2$ & 6.4 & 0 & 7.7 & $-7.7$ & 7.7 \\
$\tilde{T}^{(2)}$      &3.9 & 5.8 & 0.7 & 1.9 & 4.9 & 0.4 & 0.4 & 5.2 & 3.9 & 3.9 & 7.8 \\
$\tilde{T}^{(3)}_L$ & $-5.1$ & $-2.1+4.9 i$ & $-0.7$ &
   $2.5$ & $-4.4+5.8 i$ & $-0.6$ & $0.3$ & $-0.4+3.8 i$ & $-0.9+4.4 i$
   & $-6.2$ & 7.7 \\
$a$ &  $-0.28$ & $0.12+0.16 i$ & 0.00& 0.66 & $0.02+0.18 i$ & $-0.13$ &
   0.28 & $0.16+0.12 i$ & $0.34+0.14 i$ & $-0.33$ & 0.77\\
$a$(lQCD)     & $-0.31(2)$ &  & $0.00(1)$&  &  &  $-0.16(4)$ & & & & $-0.23(4)$&\\
 \hline\hline
    \end{tabular} 
\end{table*}
\begin{table*}[htpb]
      \renewcommand{\arraystretch}{1.6}
     \setlength{\tabcolsep}{0.1cm}
     \centering
    \caption{Decomposition of the relativistic NNLO threshold $T$-matrix elements $\tilde{T}^{(3)}_L$  (part A from diagrams Figs.~(2a-2f); part B from diagrams Figs.~(2g-2l);
part C from diagrams Fig.~3).\label{table4}}
     \begin{tabular}{c|c|cc|cc|c|ccc|cc}
     \hline\hline
$(S,I)$  & $(2,1/2)$& \multicolumn{2}{|c|}{$(1,1)$}   &\multicolumn{2}{|c|}{$(1,0)$}  &$(0,3/2)$ & \multicolumn{3}{|c|}{$(0,1/2)$}   & $(-1,1)$  & $(-1,0)$\\
Channels & $D_s K$& $D K$& $D_s \pi$ & $D K$ &$D_s \eta$ & $D \pi$ &$D\pi$  &$D\eta$ &$D_s\bar{K}$ & $D\bar{K}$ & $D\bar{K}$\\\hline
A        &   $-6.6$ & $-2.1+4.2 i$ & $-0.6$ & 0.7
    & $-3.5+6.7 i$ & $-0.8$ & $0.5$ & $-1.8+3.4 i$ & $-1.9+6.3 i$ &
   $-7.2$ & 6.7 \\
B &      1.9 & $-0.6$ & $-0.1$ & 1.9 & 0.2 & 0.2 & $-0.2$ & 0.7 & 0.7 & 1.5 & $-1.8$ \\
C & $-0.4$ & $0.7+0.8 i$ & 0.1 & 0.0 & $-1.0-0.9 i$ & 0.0 &
   0.0 & $0.7+0.5 i$ & $0.3-1.9 i$ & $-0.5$ & 2.8\\
$\tilde{T}^{(3)}_L=A+B+C$ & $-5.1$ & $-2.1+4.9 i$ & $-0.7$ &
   $2.5$ & $-4.4+5.8 i$ & $-0.6$ & $0.3$ & $-0.4+3.8 i$ & $-0.9+4.4 i$
   & $-6.2$ & 7.7 \\

 \hline\hline
    \end{tabular} 
\end{table*}

It should be pointed out that since we have neglected all the NNLO counter-term contributions, we are not in
a position to comment on the convergence behavior of either the covariant or the HM$\chi$PT results. Furthermore,
because of the nearby resonance $D_{s0}^*(2317)$, a pure $\chi$PT calculation, such as ours, in the $DK(0)$ and $D_s\eta$ channels should be taken with
care, where coupled-channel unitarity may play an important role (for a relevant discussion, see Ref.~\cite{Guo:2009ct}).

\section{Summary}
We have studied the scattering lengths of Nambu-Goldstone bosons interacting with $D$ mesons using a
covariant formulation of $\chi$PT. In particular, we have studied the recoil corrections by comparing the relativistic
with the non-relativistic results. Our studies show that the recoil corrections are
sizable, which should be kept in mind in future studies and in using the HM$\chi$PT results. Based on available information we
cannot conclude which framework is better, although in principle one should trust more covariant results, particularly when
recoil corrections are large.

Up to now, $\chi$PT describing the interactions between heavy-light mesons and Nambu-Goldstone bosons
has often been used in the non-relativistic limit. With more precise data and lattice QCD results becoming available, one may
have to study more carefully the effects of recoil corrections. The present work should be
seen as a first step in this direction.

\section{Acknowledgements}
This work was supported in part by the Alexander von Humboldt foundation
and the Fundamental Research Funds for the Central Universities (China). LSG thanks M.J. Vicente Vacas, J. Nieves, J. A. Oller, and M.F.M.  Lutz for valuable discussions and Y.R. Liu
for useful communications about the HM$\chi$PT results.

\end{document}